# Survey of Spectrum Regulation for Intelligent Transportation Systems

**Junsung Choi[1], Vuk Marojevic[2], Carl B. Dietrich[3], Jeffrey H. Reed[3], and Seungyoung Ahn[1]**

[1]The CCS Graduate school of Green Transportation, Korea Advanced Institute of Science and Technology, Daejeon 34051, Korea
[2]Electrical and Computer Engineering, Mississippi State University, Mississippi State, MS 39762, USA
[3]The Bradley Department of Electrical and Computer Engineering, Virginia Tech, Blacksburg, VA 24060, USA

Corresponding author: Seungyoung Ahn (e-mail: sahn@kaist.ac.kr)

This work was supported by Institute of Information & communications Technology Planning & Evaluation (IITP) grant funded by the Korea government(MSIT) (No.2020-0-00839,Development of Advanced Power and Signal EMC Technologies for Hyper-connected E-Vehicle). This research was supported by the MSIT(Ministry of Science and ICT), Korea, under the ITRC(Information Technology Research Center) support program(IITP-2020-2016-0-00291) supervised by the IITP (Institute for Information & communications Technology Planning & Evaluation). The work of Vuk Marojevic was supported in part by the NSF PAWR program under grant CNS-1939334.

**ABSTRACT** As 5G communication technology develops, vehicular communications that require high reliability, low latency, and massive connectivity are drawing increasing interest from those in academia and industry. Due to these developing technologies, vehicular communication is not limited to vehicle components in the forms of Vehicle-to-Vehicle (V2V) or Vehicle-to-Infrastructure (V2I) networks, but has also been extended to connect with others, such as pedestrians and cellular users. Dedicated Short-Range Communications (DSRC) is the conventional vehicular communication standard for Intelligent Transportation Systems (ITS). More recently, the 3rd Generation Partnership Project introduced Cellular-Vehicle-to-Everything (C-V2X), a competitor to DSRC. Meanwhile, the Federal Communications Commission (FCC)issued a Notice of Proposed Rulemaking (NPRM) to consider deploying Unlicensed National Information Infrastructure (U-NII)devices in the ITS band with two interference mitigation approaches: Detect-and-Vacate (DAV)and Re-channelization (Re-CH). With multiple standard options and interference mitigation approaches, numerous regulatory taxonomies can be identified and notification of relevant technical challenges issued. However, these challenges are much broader than the current and future regulatory taxonomies pursued by the different countries involved. Because their plans differ, the technical and regulatory challenges vary. This paper presents a literature survey about the technical challenges, the current and future ITS band usage plans, and the major research testbeds for the U.S., Europe, China, Korea, and Japan. This survey shows that the most likely deployment taxonomies are (1) DSRC, C-V2X, and Wi-Fi with Re-CH; (2) DSRC and C-V2X with interoperation, and (3) C-V2X only. The most difficult technical challenge is the interoperability between the Wi-Fi-like DSRC and 4G LTE-like C-V2X.

## I. INTRODUCTION

Vehicular communications technology is getting attention as the dominant technology for Intelligent Transportation Systems (ITS) in the form of Vehicle-to-Vehicle (V2V) and Vehicle-to-Infrastructure (V2I). Moreover, the possibility of connecting additional pedestrian, cellular, and Wi-Fi devices with vehicles has also been discussed. Vehicle-to-Everything (V2X) communications have the potential to enable vehicle-related safety and new services [1].

Dedicated Short-Range Communication (DSRC), standardized as IEEE 802.11p, has been the dominant protocol for vehicular communications since 1999 [2]. DSRC is now allocated spectrum in the 5.9 GHz band. Before the 3rd Generation Partnership Project (3GPP) Release 14 [3], which announced Cellular-V2X (C-V2X) for vehicular communication, DSRC was essentially the only option being considered. Many research groups such as the 5G Automotive Association (5GAA) [4], 5G Communication Automotive Research and Innovation (5GCAR) [5], Car2Car Communication Consortium (C2C-CC) [6], and Qualcomm [7] presented the advantages of C-V2X compared to DSRC. Based on [7-10], the performance of C-V2X is attractive as a substitute for DSRC in the future ITS standard. More detailed comparisons between DSRC and C-V2X are shown in Table I. Given the introduction of C-V2X, the future ITS spectrum could take one of the following turns: (1) DSRC only, (2) C-V2X only, or (3) both DSRC and C-V2X.

TABLE I
DSRC AND C-V2X SPECIFICATIONS COMPARISON

| Parameters | DSRC | C-V2X |
|---|---|---|
| Modulation and Coding Schemes | BPSK, QPSK, 16QAM, 64QAM | QPSK, 16QAM, 64QAM |
| Line coding | Convolution code | Turbo code |
| Modulation | OFDM | SC-FDMA |
| Symbol duration | 8μs | 71μs |
| Cyclic Prefix duration | 1.6μs | 4.69μs |
| Sub-carrier spacing | 156.25kHz | 15kHz |
| Transmission time | Varying according to packet length (typically 0.4ms) | Fixed to 1ms |
| Channel access mechanism | CSMA-CA | Sensing based SPS transmission |
| Timing accuracy | ±1000μs | ±0.39μs |
| Frequency accuracy | ±20ppm | ±0.1ppm |
| Multi-user allocation | Single user per symbol | Multiple users have same symbol |
| User-multiplexing | None | Possible in frequency domain |
| Re-transmission | None | Blind |
| Synchronization requirements | Asynchronous | Tight synchronization |

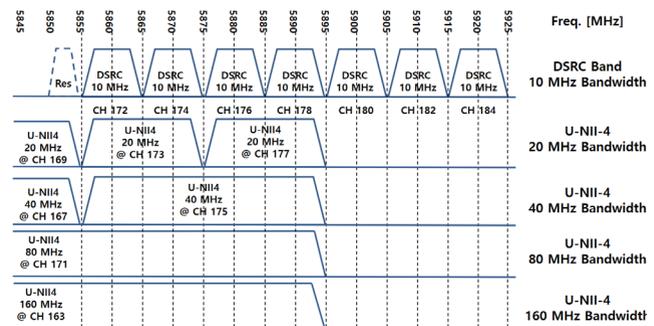

FIGURE 1. FCC's presented band plan in ET Docket 13-49

Government standardization groups completed international surveys to understand how the different countries were researching and planning to use the ITS spectrum [11-19]. However, they are yet unable to make a solid decision about which standard to choose because of the uncertainty of communication reliability, and the lack of adequate field tests. DSRC has been studied extensively and has been shown to be able to provide vehicular safety. C-V2X has the potential to interact with 5G-and-beyond cellular technologies; however, not many safety-related field trials have been conducted.

While the competition between DSRC and C-V2X was developing, a new candidate for use in the ITS band joined the discussion: Wi-Fi. The Federal Communications Commission (FCC) issued a Notice of Proposed Rulemaking (NPRM), FCC ET Docket 13-49 [20], which describes a plan for sharing the 5850–5895 MHz band between the DSRC and Wi-Fi systems to be deployed in the U-NII-4 band. The envisaged band allocation is shown in Fig 1. The interference mitigation approaches proposed are Detect-and-Vacate (DAV) and Re-channelization (Re-CH). Wi-Fi deployment in the ITS band is only being considered by the U.S. for now; however, the NPRM gives enough attention to Wi-Fi to make it a new candidate for operating in the 5.9 GHz band.

While vehicular-communication-related research for ITS is increasing [21-27], the relevant technical challenges have already been defined [28, 29]. Reference [28] presents a comprehensive survey of the 5 GHz band and [29] presents technical challenges related to V2X communications. Throughout [28, 29], researchers involved in V2X communications research have had a clear goal. However, more research with the aim of responding to the specific standard issues is needed.

With three standard options, DSRC, C-V2X, and Wi-Fi, with various interference mitigation approaches, multiple ITS spectrum usage standards can be defined. Interestingly, the U.S., Europe, Korea, and Japan selected DSRC for their ITS band, whereas China chose C-V2X. However, recently presented plans are different. For example, the U.S. is exploring the use of DSRC, C-V2X, and Wi-Fi for the band with Re-CH, whereas Europe strongly disagrees with the U.S. plan and recommends interoperable DSRC and C-V2X systems without Wi-Fi. With this variety of possible options for the standard, those shaping the standardization group in each country will need to choose a path forward. By knowing these plans, future research and selection of the technical challenges to be addressed will become more focused.

This paper reports the results of a survey of technical papers and technical challenges in response to the possible ITS band regulation taxonomies, the current regulatory plans, and major testbeds. The possible regulation taxonomies are shown in Section II. A survey of technical papers and technical challenges organized by taxonomies are presented in Section III. The ITS band-usage regulation of each country is shown in Section IV and their plans follow in Section V. Section VI presents the major testbeds in each country. Section VII summarizes the survey and Section VIII draws conclusions and presents recommendations.

## II. POSSIBLE TECHNICAL OPPORTUNITIES AND CHANLLENGES FOR THE REGULATORY TAXONOMIES BEING DEBATED

When defining the regulations for the ITS spectrum access, the specific approach will dictate the R&D priorities. If the approach is spectrum coexistence among heterogeneous radio-access technologies, the next highest priority issue would be what coexistence technique to apply. Before the FCC ET Docket 13-49 [20] was issued, there were more considerations for the regulation, but as a result of the FCC ET Docket 13-49 [20], the consideration of deploying Wi-Fi in the ITS spectrum has become the next highest priority.

The high-level option for ITS spectrum use is whether to allow single or multiple applications. The candidates for the ITS application are DSRC and C-V2X. There are arguments from 5GAA, 5GCAR, and Qualcomm that a single dominant application should use the ITS band [4, 7-10, 30-32]. Even so, using both standards for ITS has also been discussed [33-37].

If multiple standards are used for the ITS band, two operational techniques can be considered: coexistence and interoperation. For coexistence, as mentioned in [20], DAV and Re-CH are applicable for both DSRC and C-V2X. When more than one application is used in the same spectrum, DAV is applicable. The technique works because the lower priority standard vacates the band when it detects the higher-priority standard's attempt to use. The lower priority standard is only able to use the band when it is empty. Re-CH is applicable when multiple standards are deployed in different bands, on adjacent channels. The interoperation techniques considered between DSRC and C-V2X use dual interface devices, protocol conversion through a higher layer, or backhaul connection.

The deployment of Wi-Fi in the ITS band has already been presented [20]. If Wi-Fi is deployed in the band, the interference mitigation techniques could be DAV or Re-CH.

If a combination of applications is allowed to access the ITS spectrum, and considering operational techniques and the possible deployment of Wi-Fi, 15 regulatory taxonomies can be defined and classified, as shown in Fig 2.

Each taxonomy distinguishes between single and multi-standard options. For a single standard, the next consideration is whether to deploy Wi-Fi or not. The selection of appropriate interference mitigation techniques follows next. For multiple standards, the taxonomies can be specified by one of two operational techniques: coexistence or interoperation. The coexistence technologies being considered are DAV and Re-CH. The option for deployment of Wi-Fi and its mitigation technique options follow next. For the interoperability option, similar to the single standard, only the Wi-Fi deployment decision and mitigation technique selections are considered.

## III. TECHNICAL PAPERS

In this section, results of a literature survey about vehicular communication are presented. The survey is organized into the possible regulation taxonomies: DSRC only, C-V2X only, or DSRC and C-V2X. Following the survey results, the technical challenges relevant to each taxonomy are elaborated. The summary of technical papers related to a specific regulation taxonomy, shown in Fig. 2, is presented in Table II.

### A. DSRC ONLY

Reference [38] evaluates the effects of adjacent channel interference in multi-channel vehicular networks. In the model setup, a target node observes the Service Channel (SCH) 4 and various numbers of nodes transmit on SCH3 to cause adjacent channel interference on the target node. This study presents that a node tuning into a channel with low transmission power, so as to mitigate adjacent channel interference effects, would preserve the communication quality to some extent. The study also concludes that, despite the blocking, the channel-access delay might be reduced and transmissions could be less prone to collisions.

The authors in [39] analyzed the effects of adjacent channel interference levels, channel access delay, and packet loss in multi-channel vehicular networks using an adjacent channel interference model. This was done using simulations that able to control time, space, and frequency parameters for the mobile nodes. The researchers explored two scenarios: 1) vehicular nodes arranged in a square and adjacent channel interference effects measured in the center, and 2) 60 cars exponentially distributed over a 3-lane highway. Through the simulations, the effect of adjacent channel interference is significant for transmission power settings of 20 dBm and when the involved nodes are at a distance < 7 m. The results show that increased packet losses are the most evident effect of adjacent channel interference in the mixed co-channel and adjacent channel interference scenarios.

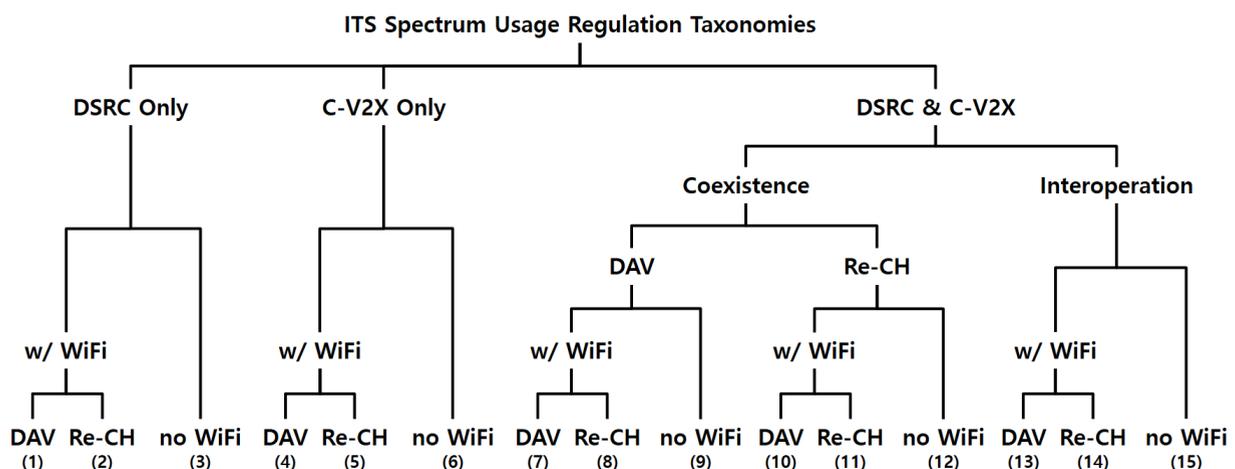

**FIGURE 2.** Possible ITS spectrum usage regulation taxonomies

Reference [40] presents analysis of the coexistence between Wi-Fi and DSRC using physical-layer challenges and MAC-layer challenges for the two systems. At short distances between the DSRC transmitter and receiver, there are no significant coexistence issues. For long-range DSRC communications, there is high DSRC packet loss due to interference from Wi-Fi, but long distances may not be as critical for safety-related DSRC applications. At medium distances, Wi-Fi outdoor would coexist better than Wi-Fi indoor; the latter creates non-negligible DSRC packet loss, which can be problematic for safety applications. The results show that even with DAV, Wi-Fi indoor can cause interference with DSRC; so reducing the Wi-Fi transmit power is recommended. The results also show that DAV provides better coexistence mechanisms for DSRC and is recommended if Wi-Fi and DSRC share the same band.

Other researchers [41] compared four dynamic spectrum sharing schemes involving Wi-Fi and DSRC: non-sharing, original sharing, Qualcomm proposal, and Cisco proposal. The Cisco proposed scheme is similar to the original sharing proposal, sharing up to channel 181 for the 20 MHz bandwidth and up to channel 177 for other bandwidths, but with band priority in DSRC. The Qualcomm-proposed scheme modifies DSRC into a 20 MHz-bandwidth signal at lower 40 MHz channels, allowing Wi-Fi to use up to channel 177, with the upper band only for DSRC. In conclusion, all of the sharing schemes provide significant improvement in Wi-Fi performance with less degraded DSRC performance. More specifically, the Cisco proposal favors DSRC while the Qualcomm proposal favors Wi-Fi.

An analysis of the impact of Wi-Fi on DSRC when both need to coexist in the same spectrum and using the Real-time Channelization Algorithm (RCA) to increase the Wi-Fi Access Point (AP) throughput performance are presented in [42]. The authors conclude that the frame aggregation with 802.11ac can severely impact DSRC performance. With DSRC set as the secondary channel and 802.11ac set as the primary channel, the increased Wi-Fi Arbitration Inter-Frame Spacing (AIFS) does not protect DSRC performance. Furthermore, they warn that certain channelization schemes should not be used because of their negative impact on DSRC performance. The proposed RCA can obtain higher average throughput than with the best static allocation scheme. By means of experiments, the authors observed that informed channelization allocation at Wi-Fi APs increases the throughput performance better than with static channel-allocation schemes.

The authors of [43] evaluated the impact of Wi-Fi transmissions on DSRC performance, using in particular, the parameters of Inter-Frame Spacing (IFS) and sensing range. They also implemented a message expiration feature in the simulations. From the simulations, the authors observed that the DSRC performance degrades with default IEEE 802.11ac parameters, IFS = 23 μs and time slot = 9 μs, and that the DSRC performance tends to be similar to the performance without Wi-Fi when the Wi-Fi IFS value is larger than 200 μs. However, when the Wi-Fi IFS value increases to 200 μs, the performance of IEEE 802.11ac decreases significantly.

A performance evaluation of DSRC with IEEE 802.11ac in adjacent channels is presented in [44]. The authors used an RF emulator for Wi-Fi with 100% channel access, and DSRC devices with varying RF attenuation. Wi-Fi of 20 MHz in channel 169 and 40 MHz in channel 167 were considered. The experiments were conducted assuming a worst-case scenario for both Wi-Fi and DSRC by controlling the channel capacity and transmission power. A noticeable effect of Wi-Fi's on the DSRC performance occurs when the Wi-Fi-to-DSRC receiver attenuation is about 10 dB lower than the DSRC transmitter-to-receiver attenuation.

Following is a list of some of the technical challenges for DSRC-only regulatory taxonomy.

- **Evaluation of adjacent channel interference:** Evaluation of the Wi-Fi interference effects to determine how close to DSRC channels Wi-Fi can be placed, how to revise the current Wi-Fi and DSRC spectral masks, and how to mitigate in-band interference to achieve the desired performance for both systems.
- **Evaluation of co-channel interference:** Evaluation of suitable guard bands between Wi-Fi and DSRC transmission to use in combination with advanced scheduling schemes to deploy both Wi-Fi and DSRC in adjacent channels.
- **Signal detection:** Technical improvement of Wi-Fi to achieve more reliable detection of DSRC for an advanced DAV algorithm with reduced vacating time, and thus a reduced transmission gap.
- **Advanced system configurations:** Exploration of current Wi-Fi standards to provide suitable parameter values for improving its coexistence with DSRC.
- **Spectrum sharing technology:** Technical improvement of the DAV algorithm, or creation of new spectrum sharing algorithms for Wi-Fi and DSRC coexisting in the same channel.

### B. C-V2X ONLY

The authors of [45] analyzed the performance of C-V2X Mode 4, which they call LTE-V. They compared the performances of DSRC and LTE-V in fast- and slow-moving vehicle environments and evaluated the LTE-V performance for different modulation schemes. From their analysis, the authors observed that LTE-V outperforms DSRC when DSRC is using a low data rate as an alternative DSRC operating mode. This is because of the improved link budget, the support for redundant transmissions per packet, and different sub-channelization schemes. However, careful configuration of parameters is needed for more efficient use.

The authors of [46] presented analytical models of C-V2X Mode 4. The average Packet Delivery Ratio (PDR) as a function of the distance between the transmitter and receiver is presented in this paper. The models were validated using variations of transmission power, packet transmission rate, modulation and coding scheme, and traffic density. Comparisons of the model and the simulation were also conducted and the models were found to be within 2.5% of the simulated results.

An analysis of the impact of configurations tuning to C-V2X in highly congested vehicular networks is presented in [47]. The selected configurations are modulation and coding scheme, distance between transmitter and receiver, probability of selecting a new resource, and reference signal received power of the transport block. The performance metrics are the Packet Error Rate (PER) and the Inter-Packet Gap. Throughout the evaluation, the importance of the configuration tuning for the best performance and reliability in highly congested networks is illustrated. However, at the same time, the authors mention that a uniform configuration should be adopted for the vehicular application perspective.

The authors of [48] showed the performance of C-V2X Mode 4 in relation to different parameters of PHY and MAC, evaluated their influence, and suggested guidelines for improvements. From the simulations, the authors conclude that modification of the PHY and MAC parameters is less effective in a low-to-medium congested network, but is significant in a highly congested network. For the PHY layer, the parameters affect the performance less, but possibly achieve improvements in sensing. For the MAC layer, modification of the resource reselection probability can provide a trade-off between a high packet-reception probability and a low update delay.

Performance evaluations of C-V2X with co-channel and adjacent channel interference from Wi-Fi are discussed in [49]. For the co-channel coexistence scenario, the authors use reduced Detect-and-Mitigate (DAM), absolute DAM, DAV, and Tiger Team Sense and Vacate mechanisms. From the simulations, the authors observe that the complete vacation of Wi-Fi must be done to get good C-V2X performance and large contention configurations can minimize the impact of Wi-Fi on C-V2X. For the adjacent channel coexistence scenario, the authors use Wi-Fi interference sources as similar characteristics of U-NII-3, 4, and 5. From U-NII-3 simulations, the authors observed that channel 155 must be avoided. From U-NII-4 simulations, the authors observed that many adjacent channels do not affect C-V2X performance, except on channel 180. Moreover, when C-V2X is deployed in channel 182 or 184, the deployment of Wi-Fi in channels 171 and 175 must be avoided. From U-NII-5 simulations, all the channel deployments in channel 189, 191, and 195 should be restricted or prohibited from use.

We conclude that the important technical challenges for the C-V2X-only regulatory taxonomy are as follows.

- **V2X performance:** Analysis of the C-V2X performance with responses to DSRC, and fair comparison between C-V2X and DSRC, as well as whether which standard is more suitable than the other for ITS.
- **Adjacent channel interference:** Evaluation of adjacent-channel interference stemming from Wi-Fi to determine how close Wi-Fi can be placed to the C-V2X channels, how to revise the current Wi-Fi and C-V2X spectral masks, and how to avoid the in-band interference stemming from C-V2X to achieve the desired performance of both systems.
- **Co-channel interference:** Evaluation of co-channel Wi-Fi interference effects to determine the necessary geographic spacing between Wi-Fi and C-V2X, or definition of the appropriate exclusion zones, in combination with advanced scheduling schemes for time-division access and interference-free co-channel coexistence.
- **Signal detection:** Technical improvement of Wi-Fi to detect a C-V2X signal with higher reliability by advancing the DAV algorithm, or developing new algorithms, and achieve shorter vacating-time transmission gaps.
- **Advanced system configurations:** Exploration of suitable parameter values in the current Wi-Fi standards to achieve the best coexistence performance with Wi-Fi and C-V2X.
- **Spectrum sharing technology:** Technical improvement of the spectrum-sharing algorithms for Wi-Fi and C-V2X systems transmitting in the same channels, as alternatives to the DAV algorithm.
- **C-V2X scheduling and congestion control:** Technical improvements of the scheduling and congestion control mechanisms for C-V2X networks beyond what is standardized.

### C. DSRC AND C-V2X

An efficient Cooperative Awareness Message (CAM) forwarding mechanism, or message relaying, for the coexistence between C-V2X and DSRC is proposed in [50]. A Quality of service-aware Relaying algorithm (QR) is proposed with the functionality of mitigating channel congestion to reduce unnecessary CAM relays. The QR algorithm was applied to the LTEV2Vsim with IEEE 802.11p and LTE-V2V MAC layers. The results show that the QR algorithm can provide performance gains for dual interface vehicles that can communicate with C-V2X and DSRC.

The authors of [51] proposed a solution to achieve interoperability between DSRC and C-V2X. The interoperability solutions are divided into multi-access issues and multi-operator issues. For multiple access, the suggested solution is to use straightforward conversion between DSRC and C-V2X packets. The paper reports that the higher layer protocol stacks perform identically. For a multi-operator system, the suggested solution is to use a Multi-access Edge Computing infrastructure for the role of bridge, to transcode messages between DSRC and C-V2X.

A cognitive protocol converter as a solution for interoperability between DSRC and LTE is proposed in [52]. The converter identifies the standard received packet and converts the data format into the desired standard. The authors propose the conversion process as a knowledge-based updating process.

We conclude this section with a list of the key technical challenges of the DSRC and C-V2X coexistence taxonomy.

- **Adjacent channel interference:** Evaluation of the adjacent interference effects between Wi-Fi and DSRC and Wi-Fi and C-V2X to determine how close Wi-Fi can be placed to DSRC or C-V2X channels, how to revise the current Wi-Fi and DSRC or C-V2X spectral masks, and how to mitigate the in-band interference between Wi-Fi and DSRC or C-V2X to achieve the desired performance of all three systems.
- **Co-channel interference:** Evaluation of the co-channel effects of interference between DSRC and C-V2X for appropriate frequency spacing between Wi-Fi and DSRC to reduce negative effects on each other, and advance of the scheduling scheme to deploy Wi-Fi, DSRC, and C-V2X.
- **Interoperability methodology and architecture:** Methodology and technical architecture for enabling the interoperability between DSRC and C-V2X, especially in the vehicle-to-vehicle communication mode.
- **Backhaul compatibility:** Shared or interoperable backhauls for DSRC and C-V2X for upper-layer sharing methods, multi-standard conversion methods, and reliable backhaul connection.
- **Signal detection at Wi-Fi node:** Technical improvement of Wi-Fi to detect a DSRC or C-V2X transmission for use with the advanced DAV algorithm to reduce the mutual effects and decrease the vacating time to reduce the Wi-Fi and DSRC or C-V2X transmission gap.
- **Signal detection at DSRC and C-V2X nodes:** Technical improvement for the detection and identification of DSRC by C-V2X and vice-versa for decreasing the vacating-algorithm process time and increasing accuracy of the algorithm.
- **Advanced interference management:** General improvements of performance through advanced channelization, interference avoidance, and mitigation.

TABLE II
SUMMARY OF TECHNICAL PAPERS

| Application Usage | Reference | Taxonomy Index | Contribution |
|---|---|---|---|
| DSRC Only | [38] | (2) | · Node tuning into a channel with a low transmission power to mitigate adjacent channel interference effects. |
| | [39] | (2) | · Effect of adjacent channel interference is significant in 3-lane highway environment with transmission power settings of 20dBm. |
| | [40] | (1) | · DAV provides better DSRC performance in coexistence scenario.<br>· Even with DAV, Wi-Fi can cause interference to DSRC. |
| | [41] | (1) | · Four dynamic spectrum sharing schemes are conducted: non-sharing, original sharing, Qualcomm's, and Cisco's proposed sharing.<br>· All sharing schemes provide significant improvement than non-sharing.<br>· Cisco's proposal is a favor to DSRC while Qualcomm's proposal is for Wi-Fi. |
| | [42] | (1) | · 802.11ac's frame aggregation can severely impact the DSRC performance.<br>· Informed channelization allocation at Wi-Fi APs can increase the throughput. |
| | [43] | (1) | · DSRC performance degrades with default IEEE 802.11ac parameters, IFS = 23μs and time slot = 9μs.<br>· When IFS value increases to 200μs, DSRC tends to similar performance as operating without Wi-Fi scenario. |
| | [44] | (2) | · Noticeable effect is observed when Wi-Fi to DSRC receiver attenuation is about 10dB lower than DSRC transmitter to receiver attenuation. |
| C-V2X Only | [45] | (6) | · LTE-V outperforms DSRC when DSRC is using a low data rate. |
| | [46] | (6) | · Analytical models of C-V2X Mode4 is presented; the model is 2.5% within simulated results. |
| | [47] | (6) | · PER and Inter-Packet Gap performances are analyzed with response to modulation and coding scheme, distances between transmitter and receiver, probability of selecting a new resource, and reference signal received power.<br>· A uniform configuration should be adopted for vehicular application perspective. |
| | [48] | (6) | · PHY layer parameters affect the performance less, but possibly achieve improvements in sensing.<br>· Modification of the resource reselection probability can provide a trade-off between a high packet reception probability and a low update delay. |
| | [49] | (4), (5) | · DAM, absolute DAM, DAV, and Tiger Team Sense & Vacate mechanisms are analyzed.<br>· In co-channel coexistence scenario, complete vacation of Wi-Fi and large contention configurations can reduce Wi-Fi to C-V2X interference effect.<br>· Channel 155 must be avoided for U-NII3, no effect except on channel 180 for U-NII4, and all channel deployments are restricted or prohibited for U-NII5. |
| DSRC and C-V2X | [50] | (9) | · Proposed an algorithm to mitigating channel congestion to reduce unnecessary CAM relays. |
| | [51] | (15) | · Straightforward conversion between DSRC and C-V2X packets for multi-access issue.<br>· Multi-access Edge Computing infrastructure as role to bridge to transcode messages for multi-operator issue. |
| | [52] | (15) | · A cognitive protocol converter to converts the data format into the desired standard. |

## IV. CURRENT SPECTRUM REGULATION

The current ITS band regulations for several countries are presented in the following section. As shown in Fig. 3, the ITS spectrum allocations differ from country to country and deviate from the International Telecommunication Union Radiocommunication Sector (ITU-R) recommendation, 5725–5875 MHz [53].

### A. U.S.

The FCC, in its Report and Order FCC-99-305 [54], initially allocated 75 MHz in the 5.9 GHz band for vehicular communications in 1999. Amended allocations were presented in 2004 and 2006 [55, 56]. In 2016, FCC refreshed the record [20]. The FCC selected the frequency range 5850–5925 MHz. Each channel is 10 MHz wide as per the recommendation of ITS, with 5 MHz reserved at the lower end of the band. Channel 178 is to be used as the Control Channel (CCH); channels 174, 176, 180, and 182 for SCH non-safety applications; and channels 172 and 184 for SCH safety applications.

### B. EUROPE

The European Commission issued the Commission Decision 2008/671/EC [57], which legally forces 5875–5905 MHz to be used for traffic safety-related applications in the European Union. The European Conference of Postal and Telecommunications Administrations (CEPT) harmonization was applied by the European Communications Committee (ECC) Decision and indicates using 5905–5925 MHz for an extension of the ITS spectrum. The ECC Recommendation (08)01 [58] suggests that CEPT administrations use the 5855–5875 MHz band for non-safety applications. Based on these recommendations, ETSI issued the standard EN 302 571 [59] in which it defines the requirement that ITS equipment operate at 5855–5925 MHz. The current spectrum allocation was designed following the technical recommendations TR 102 492-1 [60] and TR 102 492-2 [61], where 5855–5875 MHz is assigned for non-safety related applications, 5875–5885 MHz for road-safety and traffic-efficiency applications, 5885–5905 MHz for critical road-safety applications, and 5905–5925 MHz for road-safety and traffic-efficiency applications. More specific ITS application operation scenarios are described in [62-65].

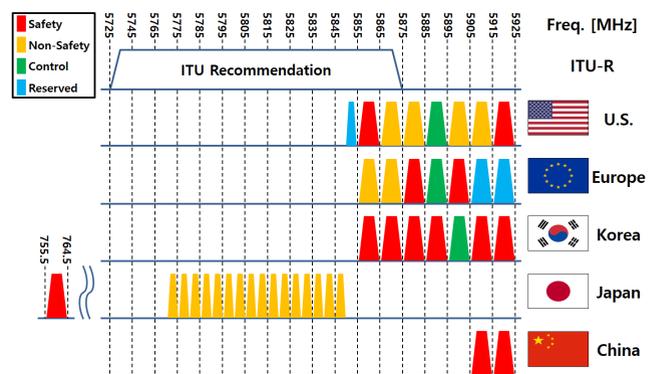

FIGURE 3. ITS spectrum usages for U.S., Europe, Korea, Japan, and China

### C. CHINA

China used 20 MHz in the 5.9 GHz band, with 5905–5925 MHz for ITS [17, 66]. In 2018, the Bureau of Radio Regulation (BRR) nominated the State Radio Regulation of China (SRRC) and the Telematics Industry Application Alliance (TIAA) to lead a study of ITS spectrum policy [66-68]. Different from other countries, China considers C-V2X the only option for ITS safety applications. As a result of research and trials, the Chinese Ministry of Industry and Information Technology (MIIT) regulated the band in October 2018 for the Internet of Vehicles (IoV), based on LTE-V2X technology [66-68].

### D. KOREA

The Telecommunications Technology Association (TTA) documented DSRC communications between Road-Side Units (RSU) and On-Board Units (OBU) in 2006 for the 5.8 GHz band [69], as well as test standards in 2007 [70]. More specific standards are shown in [71-74]. After several trials, the Ministry of Science and Information and Communications Technology (MSIT) allocated 5855–5925 MHz for C-ITS applications in 2016 [66]. The spectrum is divided into seven different channels of 10MHz each, where the fifth channel, 5895–5905 MHz, is meant for CCH and the others for SCH.

### E. JAPAN

Japan allocated two ITS applications—ITS connect and ETC/ETC2.0—in two different bands. ITS connect is similar to ITS communication regulations in the U.S., Europe, and Korea for the 5.9 GHz band, but it is allocated at 755.5–764.5 MHz with only one channel available [66, 75]. Despite the different frequencies, the structure is the same as for DSRC [76-78]. ETC/ETC2.0 is defined for tolling applications and is allocated to 5770–5850 MHz with fourteen channels: seven downlink and seven uplink channels of 5 MHz bandwidth each [66].

## V. FUTURE REGULATORY PLANS

The current ITS spectrum usage regulations are similar for many countries, and most of them rely on DSRC. However, their future plans for the ITS band differ significantly from the current status. These plans are presented below.

### A. U.S.

The FCC issued a NPRM regarding potential use of the 5.9 GHz band for U-NII devices. According to the FCC Docket ET 13-49 [20], the FCC is considering sharing the 5850–5895 MHz band between DSRC and U-NII devices. The primary unlicensed devices considered in the FCC NPRM use a signal based on IEEE 802.11ac that operates in the U-NII-4 band. In December 2019, FCC voted to grant the lower 45MHz channels for U-NII-4 devices [79]. Furthermore, according to the FCC Docket ET 19-138 [80], FCC proposed to revise the ITS rule to permit C-V2X at 5905–5925 MHz; suggesting that only 10 MHz (5895–5905 MHz) be used for DSRC.

The channel allocations being considered for DSRC, C-V2X, and U-NII-4 are shown in Fig. 4. In the FCC NPRM, two interference mitigation approaches are presented: DAV and Re-CH. DAV represents no changes to DSRC. It requires unlicensed devices to vacate the channel to avoid interfering with the DSRC signal by detecting the DSRC signal in channel 178, or lower channels. Re-CH is a re-allocation process whereby safety-related DSRC applications use the upper 30 MHz (channels 180, 182, and 184) while non-safety-related DSRC and U-NII devices share the lower 45 MHz (channels 172, 174, 176, and 178).

### B. EUROPE

Europe chose ITS-G5 as the dominant standard for the ITS spectrum and services. However, the European Council (EUCO) rejected the Delegation Act, which set ITS-G5 as the only application for C-ITS, in 2019 [81]. With the stated position of opposition by EUCO, Europe may still be technology-neutral, but seems to favor C-V2X [82-84].

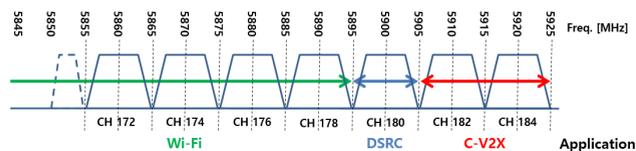

**FIGURE 4.** FCC proposal for Wi-Fi, DSRC, and C-V2X in ITS band

Even though the European position is considerably technology-neutral, they are strongly opposed to the FCC plan to segment and segregate the spectrum [80]. They also do not support sharing the spectrum with other applications such as Wi-Fi [85]. Europe's stand is technology-neutral for the ITS application and establishes interoperability and backhaul compatibility with ITS-G5 as the highest priority for regulatory decisions [85].

### C. CHINA

After regulating the ITS spectrum for C-V2X in 2018, there has been no further public announcement related to spectrum regulation in China. China still has C-V2X as its only option and plans to extend the C-V2X trials [66-68].

### D. KOREA

The "ITS-master plan" was established in 2012 and the discussions on using DSRC also started [86, 87]. Most of the previous research was done considering DSRC as the only deployment option [13]. However, MSIT announced a plan to provide C-V2X experimental services in September 2019 [88-90]. MSIT has encouraged more research about C-V2X and other ITS technologies in order to make the best regulation decisions. The relevant ITS standard decision-makers, MSIT and the Ministry of Land, Infrastructure, and Transport (MOLIT), have not provided a definite policy decision yet, but they are presently open to both DSRC and C-V2X.

### E. JAPAN

The Ministry of Internet Affairs and Communications (MIC) reported a connected-car plan in 2017 [91]. The prime ministry presented a roadmap for ITS in 2018 [92]. In both reports, the priority of ITS applications increased significantly; however, a specific technology to use was not mentioned. Most likely, they will follow the global trend, which is as yet undefined [91].

## VI. TESTBEDS

The various national standardization groups pushed efforts in the forms of research and testbeds with specific safety services to be able to make their own informed decisions for ITS spectrum-usage regulation. The major testbeds are presented in this section.

### A. U.S.

The U.S. Department of Transportation (U.S. DoT) is leading the ITS testbeds and trials across the U.S. It cooperates with several cities and universities to create ITS testbeds. New York City, Tampa, Florida, and Wyoming have been selected as the national testbed implementation areas [93]. Moreover, universities and research groups in Michigan and Virginia cooperate with the U.S. DoT to develop areas for testing. In this section, the U.S. DoT plans and testbeds for ITS application are discussed. Table III summarizes the ITS applications for each testbed.

### 1) U.S. DOT TEST PLAN

The U.S. DoT is evaluating the feasibility of spectrum sharing between DSRC and UNII devices with the objective to determine whether DSRC can provide safety-critical messages while U-NII-4 devices operate as interference sources [94]. The tests consider three types of interference: at the DSRC receiver, at the transmitter, and adjacent/non-adjacent channel interference. The spectrum sharing techniques are DAV and Re-CH.

The common performance metrics are PER, throughput, latency, and jitter. For testing the Re-CH scenario, the performance metrics are detection threshold, which is the probability of detecting DSRC, Packet Completion Rate (PCR), which corresponds to the number of successfully received packets, and the ratio between the packets in the queue and the successfully transmitted packets. They also include the Inter Arrival Time (IAT), which is the time between two received packets and the Inter Departure Time (IDT), which is the time between two transmitted packets. For testing the DAV scenario, the performance metrics are detection threshold, PCR, IAT, IDT, and channel-move time: the time between when the DSRC preamble is detected and when the IEEE 802.11 transmission has started.

The U.S. DoT is also testing LTE-C-V2X operation and performance for V2V/V2I safety applications. The types of tests are interference, scalability, interoperability, system dynamics and congestion, and validation.

### 2) NEW YORK DOT (NYCDOT) PILOT

The NYCDOT pilot project [95, 96] has the objective to improve the safety of travelers and pedestrians through V2V, V2I, and Infrastructure-to-Pedestrian (I2P) communications. The pilot project is primarily focused on safety-related applications. The test areas include three places: a 4-mile segment of Franklin D. Roosevelt Drive in Manhattan, four one-way corridors in Manhattan, and a 1.6-mile segment of Flatbush Avenue in Brooklyn. The NYCDOT pilot uses DSRC and deploys approximately 300 RSUs across the testing areas.

### 3) TAMPA-HILLSBOROUGH EXPRESSWAY AUTHORITY (THEA) DOT PILOT

The THEA pilot project [97, 98] defines as its objectives to relieve congestion, reduce collisions, and prevent wrong way entry. The pilot test area is deployed in downtown Tampa. The pilot uses DSRC with approximately 1,000 vehicles, 10 buses, 8 trolleys, and 47 RSUs.

### 4) WYOMING DOT (WYDOT) PILOT

The WYDOT pilot project [99, 100] defines its main objective as reducing the number of harsh weather-related accidents. The pilot area, I-80, usually is subjected to blowing snow during winter, and fog and high winds during summer; therefore, DOT selected the area as the best place to test weather-related applications. The project uses DSRC and the testing involves approximately 400 OBU-equipped vehicles, 150 heavy trucks, and 75 RSUs. WYDOT is unique in that it provides collected travel information through the Wyoming 511 app and a Commercial Vehicle Operator Portal (CVOP).

### 5) ANN ARBOR CONNECTED VEHICLE TEST

TABLE III
U.S. DoT TESTBEDS' PROVIDING ITS SERVICES

| Safety Category | ITS APPLICATIONS | DoT Pilot Project | | |
|---|---|---|---|---|
| | | NYC | THEA | WY |
| V2I Safety | Curve Speed Compliance | O | | |
| | Distress Notification | | | O |
| | Emergency Communications and Evacuation Information | O | | |
| | End of Ramp Deceleration Warning | | O | |
| | I2V Situational Awareness | | | O |
| | Oversize Vehicle Compliance | O | | |
| | Pedestrian Collision Warning | | O | |
| | Red Light Violation Warning | O | | |
| | Speed Compliance | O | | |
| | Speed Compliance/Work Zone | O | | |
| | Spot Weather Impact Warning | | | O |
| | Work Zone Warning | | | O |
| | Wrong Way Entry | | O | |
| V2V Safety | Blind Spot Warning | O | | |
| | Emergency Electronics Brake Lights | O | O | |
| | Forward Collision Warning | O | O | O |
| | Intersection Movement Assist | O | | |
| | Lane Change Warning/Assist | O | | |
| | Vehicle Turning Right in Front of a Transit Vehicle | O | O | |
| Pedestrian | Mobile Accessible Pedestrian Signal System | O | | |
| | Pedestrian in Signalized Crosswalk | O | | |
| Mobility | Intelligent Traffic Signal System | O | O | |
| | Transit Signal Priority | | O | |

*AACVTE and VCC are not included due to not enough information.

ENVIRONMENT (AACVTE)

AACVTE [101, 102] is an expanded project from the Safety Pilot Model Deployment (SPMD), which is a $30 million research project funded by the University of Michigan Transportation Research Institute (UMTRI) and U.S. DoT. The project launched in 2012 and an upgrade started in 2015. The project is located in the northeast quadrant of the City of Ann Arbor (Michigan), and has as its goal to develop the world's largest operational and real deployment area for connected vehicles. This project is also about the transition from research mode to operational deployment, and from a government-funded project to a self-sustainable project. The project is currently deployed over 73 lane miles, using 75 RSUs and over 2,500 connected vehicles. They are currently upgrading to include 5,000 vehicles, 45 street locations, and 12 freeway sites.

6) VIRGINIA CONNECTED CORRIDOR (VCC)

The Virginia Department of Transportation (VDoT), Virginia Tech Transportation Institute (VTTI), University of Virginia, and Morgan State University have partnered and initiated the VCC with the objective of integrated connectivity within the transportation systems [103]. The testbeds are located on VTTI's Smart Road, I-66, I-495, U.S.-29, and U.S.-50 with more than 60 RSUs. These are connected through a backhaul network via DSRC and cellular communications. The VCC provides an open application development environment, so that third-party developers can minimize time to deploy, test, and demonstrate their applications. Developers can generate the applications directly through the VCC cloud computing environment or VCC public Application Programming Interface (API). The future plans for VCC are not presented; they seek a third party that is willing to test and further develop their platform.

B. EUROPE

European countries are cooperating on tests and trials for ITS development. They selected a common platform, the C-Roads platform, to co-develop the service by sharing what they observe. Safety applications in Europe are categorized in three phases: Day 1, Day 2, and Day 3 services. Day 1 services mainly focus on exchanging information for enhancing cautious driving. Day 2 services focus on improving quality of Day 1 services and sharing awareness information. Day 3 services add more services such as sharing intentions and supporting negotiation and cooperation. Most of the European testbeds focus on Day 1 services. This section introduces some of the common European vehicular-communication research platforms and research groups. Table IV summarizes the supported safety applications research.

1) C-ROADS PLATFORM

The C-Roads Platform [104] is a joint initiative of 16 European countries. Through cooperation, the deployment of harmonized and interoperable C-ITS services across European countries is possible. Within the participating countries, they are sharing experiences and knowledge about deployment, implementation issues, and user acceptance. The European C-ITS services can evolve together and possibly achieve transnational interoperability, as well as ensuring European cohesion in the deployment of C-ITS.

2) C-ITS CORRIDOR

The C-ITS Corridor [104, 105] is a joint development between The Netherlands, Germany, and Austria; which started in 2016. The project was initiated for R&D and evaluations of Field Operational Tests (FOTs) for Road Works Warning (RWW) and vehicle data for traffic management services. There are plans to extend the C-ITS Corridor for evaluating the interoperability of European C-ITS solutions using Wireless Access in Vehicular Environment (WAVE) and cellular networks.

3) SCOOP@F

The SCOOP@F [104, 106] is the French C-ITS pilot deployment project currently installed at five sites: Ilde-de-France, the "East Corridor" between Paris and Strasbourg, Brittany, Bordeaux, and Isere. Approximately 3,000 vehicles can be connected along 2,000 km of roads. SCOOP@F was the first flagship C-ITS project in Europe. Its goal was to improve road safety for workers and the validation of C-ITS services with a hybrid communication solution. The project concluded in 2019 with on-site demonstrations of hybrid solutions and the C-ITS architecture and system security. Although the project ended, work continues through other European C-ITS projects.

4) A2/M2 CONNECTED VEHICLE CORRIDOR (A2/M2 CVC)

The A2/M2 CVC [104, 107] was initiated to test the infrastructure, data management, and service delivery necessary for connected vehicles. It was implemented across approximately 100 km of United Kingdom roads: a trunk road (A2), motorway network (M25 and M2), and Kent local roads (A229/A249). The A2/M2 CVC is expected to deliver the functional and technical specifications to ensure future UK deployment services through both ITS-G5 and Cellular, as well as "Hybrid" communication solutions.

5) NORDICWAY / NORDICWAY2

The NordicWay and NordicWay2 [104, 108, 109] are C-ITS pilot projects running in the Nordic countries: Denmark, Finland, Norway, and Sweden. The NordicWay is a three-year long pilot project (2015~2017) and NordicWay2 is the follow-up project. Both projects were initiated to test and demonstrate the interoperability of a cellular system for Day-1 C-ITS services, to evaluate enhanced traffic safety, and to support the infrastructure readiness. These projects are of particular interest because of the frequent ability to test under snowy and icy arctic conditions.

6) INTEROPERABLE CORRIDORS (INTERCOR)

InterCor [104, 110, 111] is a project to connect the C-ITS initiatives of the C-ITS Corridor [105], SCOOP@F [106], A2/M2 CVC [107], and the Belgian C-ITS initiatives [104];

This article has been accepted for publication in a future issue of this journal, but has not been fully edited. Content may change prior to final publication. Citation information: DOI 10.1109/ACCESS.2020.3012788, IEEE AccessTABLE IV
EUROPE TESTBEDS' PROVIDING DAY-1 APPLICATIONS

| | PILOT PROJECT | | | | | | | | | | |
|---|---|---|---|---|---|---|---|---|---|---|---|
| | C-ITS Corridor | SCOOP@F | A2/M2 CVC | NordicWay/ NordicWay2 | Belgium Pilot | Czech Pilot | Slovenian Pilot | Hungarian Pilot | Italian Pilot | Portuguese Pilot | Spanish Pilot |
| INVOLVED COUNTRIES | AT, DE, DL | FR | UK | DK, FL, NO, SE | BE | CZ | SL | HU | IT | PT | ES |
| STANDARD | ETSI G5, Cellular | ETSI G5, Cellular, WiFi | ETSI G5, Cellular | Cellular | ETSI G5, Cellular | ETSI G5, Cellular, WiFi | ETSI G5, Cellular | ETSI G5, Cellular | ETSI G5, Cellular | ETSI G5, Cellular, DATEX II | ETSI G5, Cellular |
| Emergency electronic brake light | | O | | O | | O | | | O | O | O |
| Emergency vehicle approaching | O | O | | O | | O | | | | O | O |
| Slow or stationary vehicle | O | O | | O | O | O | O | | O | O | O |
| Traffic jam ahead warning | O | O | | O | O | O | O | O | O | O | O |
| Hazardous location notification | O | O | | O | O | O | O | O | | O | O |
| Road works warning | O | O | O | O | O | O | O | O | O | O | O |
| Weather conditions | O | O | | O | O | O | O | O | O | O | O |
| In-vehicle signage | O | O | O | | O | O | O | O | O | O | O |
| In-vehicle speed limits | O | | | O | O | O | O | O | O | O | O |
| Probe vehicle data | O | O | O | O | | O | | O | O | O | O |
| Shockwave damping | O | | | | O | | | | | O | O |
| Green Light Optimal Speed Advisory (GLOSA) | O | O | O | O | | | O | O | | O | O |
| Time To Green (TTG) | O | O | O | O | | | O | O | | O | O |
| Signal Violation | | | | O | | O | | O | | O | O |
| Intersection safety | | | | O | | O | | O | | O | O |
| Traffic signal priority request by designated vehicles | | | | O | | O | | | | O | O |

AT: AUSTRIA, BE: BELGIUM, CZ: CZECH REPUBLIC, DE: GERMANY, DL: NETHERLANDS, DK: DENMARK, ES: SPAIN, FL: FINLAND, FR: FRANCE, HU: HUNGARY, IT: ITALY, NO: NORWAY, PT: PORTUGAL, SE: SWEDEN, SL: SLOVENIA, UK: UNITED KINGDOM

and makes a connection to the C-Roads platform. The project plans to provide interoperable C-ITS services by the European corridor network and a testbed for Day-1 C-ITS service development and deployment. The intended contributions of this project included upgraded specifications for ITS-G5, hybrid communications, Public Key Infrastructure (PKI), and security protocols and guidelines for future pilot operations. The project ended in 2019 and achieved technical evaluations of interoperability and availability of services across countries. It was found that the way information is presented to drivers has a significant impact on safety, and concluded that integration of existing navigation applications or devices is important. The project demonstrated successful interoperability among the involved countries. More work is still needed to integrate the ITS-G5 in a common European data framework.

VOLUME XX, 2020 11This work is licensed under a Creative Commons Attribution 4.0 License. For more information, see https://creativecommons.org/licenses/by/4.0/.

## C. CHINA

In November 2016, BRR of MIIT set C-V2X trials in six cities in China: Beijing, Shanghai, Chongqing, Changchun, Wuhan, and Hangzhou [16, 66-68]. Of the six trials, a few unique trials are presented in this section.

### 1) SHANGHAI PILOT

Shanghai International Automobile City, in cooperation with SAIC motor and NIO companies, constructed a pilot area to test smart cars and V2X network communications [16, 67, 68, 112]. The test area is about 5.6 km of a public road in the Jiading District. The testing features include transmitting and evaluating information about speed limit, traffic light identification, pedestrian and non-motor vehicle identification, and lane keeping.

### 2) INTELLIGENT VEHICLES INTEGRATED TEST AREA (I-VISTA)

The platform i-VISTA, is for C-V2X and was implemented in Chongqing, China [16, 66-68, 113]. The platform is conducted by the China Automotive Engineering Research Institute (CAERI). Fifty test scenarios are available for this platform. The unique features of i-VISTA are the terrain and interchanges. Chongqing is a city in the mountains that has high-speed loops, long tunnels, ramps, bends, bridges, and avenues. Moreover, the road network in the city is layered and may be the most complex road interchange system in China.

### 3) WUXI C-V2X PROJECT

The Wuxi C-V2X project is a city-level C-V2X pilot project deployed in Wuxi, China [16, 66-68, 114] and running in collaboration with private and public partners. The pilot project area is 170 km$^2$ with more than 240 implemented infrastructure. The project includes more than 20 V2V, V2I, and Vehicle-to-Pedestrian (V2P) scenarios to test.

## D. KOREA

### 1) U-TRANSPORTATION

The u-Transportation project was active between 2006 and 2012 [13]. The main contribution of the project was developing the core technology for ITS services through V2V, V2I, and sensing systems. The project was divided into three different tasks: technology development, service development, and system engineering. With the achieved advances in technology, the following services were enabled in Korea: Safety, Efficiency, and Environment (SEE)-advisor; express entering notification service, non-signal intersection guidance service, V2X based warning service, the u-Transportation based ITS monitoring service, bird-eye view service, follow-me service, and virtual Video Management Systems (VMS) service.

### 2) SAFETY/SUSTAINABLE, MOBILITY, ADVANCED AND AUTOMATION, RELIABILITY, TOMORROW (SMART) HIGHWAY

The SMART highway project started in 2008 with the goal of implementing C-ITS in practical settings [13] by combining IT, communications technology, vehicle technology, and road technology. The project concluded in 2014 with basic V2X communications technology and functionality. The project was divided into four tasks: road infrastructure technology, road-IT-based traffic operation technology, road-vehicle connection technology, and a study of applications for testbeds. As a result of the project, the following services emerged: event-share service, multilane smart-tolling service, road information-based vehicle control service, obstacle identification service, emergency warning service, chain-reaction collision avoidance service, virtual VMS service, and the V2I and V2V WAVE communication service.

### 3) C-ITS PILOT PROJECT

The C-ITS pilot project was initiated in 2014 and finished in 2017. It was intended to achieve service stability based on the technologies achieved in the u-Transportation and SMART Highway projects [13, 66]. For initial testing, the project selected public services: collision avoidance, road condition and weather information, work-area notification, intersection collision avoidance, yellow bus operation, school zone/silver zone warning, emergency warning, emergency vehicle aid, location-based vehicle data collection, location-based traffic information, smart tolling and public traffic-control services. The test sites were on expressways, national and urban roads in the cities of Daejeon and Sejong, and included 87.8 km [13, 66]. Another C-ITS implementation project started in 2018 in the Seoul metropolitan ring road and Kyungbu express road, with a total of 128 km [59]. Korea's goal is to deploy communications services that will enable the reduction of collision accidents to zero by 2030.

## E. JAPAN

### 1) ELECTRONIC TOLL COLLECTION 2.0 (ETC2.0) SERVICE

The ETC2.0 service is the first ITS service of the MLIT "Smartway" plan, a V2I cooperative system enabled by collaboration between academia and industry, initiated in 2011 [66, 91]. "ETC2.0" was introduced to develop cooperative ITS by providing services such as ETC, navigation systems, the Vehicle Information and Communication System (VICS), dynamic route guidance, and safe driving assistance. Internet access at expressways, cashless payments, tourist information, and logistics operation support are envisaged as future support services. More than 60 million cars had ETC devices installed and more than 1,600 RSUs were installed nationwide on expressways. These RSUs are spread along 390,000 km of expressways, which correspond to almost 33% of all the public roads in Japan.

## VII. CONCLUSIONS

Vehicular communications for ITS are gaining increasing attention from industry and academia. With the development of 5G technologies, vehicular communications are no longer limited to the vehicle environment only, but rather could connect everything. The conventional standard for vehicular communication is DSRC; however, since the 3GPP Rel. 14 [3] introduced C-V2X, the latter has become seriously considered as the standard for ITS services. Furthermore, due to the FCC NPRM issued for the 5.9 GHz band, Wi-Fi has become another potential player for the ITS band.

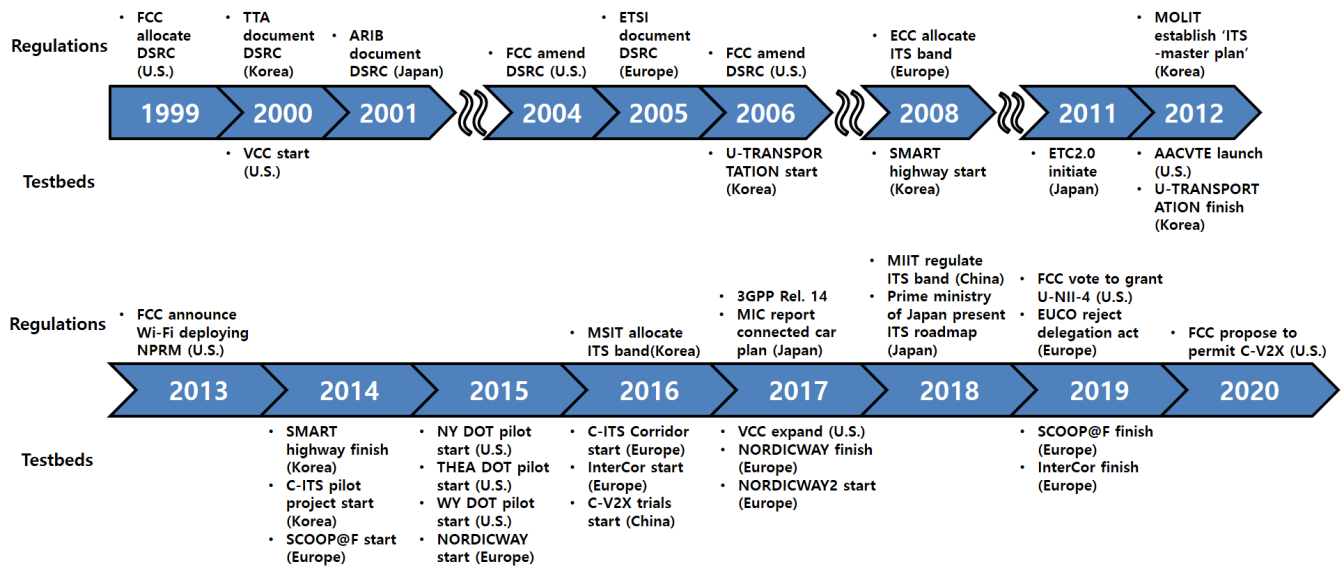

FIGURE 5. Timetable of regulation and testbed actions

With the current variety of standards and interference mitigation approaches, the key regulation plans and related technical challenges have to be identified. Because the regulation plans of different governmental standardization groups vary, so do the technical challenges vary. In this paper, we have presented the current regulations for several countries, elaborated possible regulatory options, and categorized the technical challenges and ongoing research.

The papers surveyed and the current and future regulatory plans for the U.S., Europe, China, Korea, and Japan are sorted into proper regulation taxonomies, presented in Fig. 2. The timetable of surveyed regulatory actions and testbeds is also shown in Fig. 5.

Most of the technical contributions are for the 'DSRC-only' taxonomy, and the least are for 'DSRC and C-V2X'. Interestingly, not many spectrum coexistence studies involving DSRC, C-V2X, and Wi-Fi are publicly available. Moreover, for the 'DSRC and C-V2X' coexistence papers, interoperability through backhaul is the most studied item.

We observed that currently, the U.S., Europe, Korea, and Japan have chosen DSRC for ITS applications, whereas China has chosen C-V2X. Even though the current regulations are similar, future plans are different. The U.S. is planning to deploy Wi-Fi, DSRC, and C-V2X in the ITS band with the Re-CH approach. Europe is planning to deploy interoperating DSRC and C-V2X, Korea and Japan are technology-neutral. China continues investing in C-V2X.

Because different regions have different topologies, careful technical comparisons are required. Due to different technological flavors, the research and testbeds are diverse.

By cooperating and sharing diverse technical results, the vehicular communication regulation/standardization can be effectively advanced. Global testbeds with international participation and private-public partnerships, which are recommended for technology convergence while encouraging diversity and multi-systems, may be the next big thing.

The at-scale vehicular testbeds provide different services and ITS standards. However, they have the common aim of safety of ITS users by creating a number of environmental conditions and using real environments.

We find noteworthy that the research is mainly focused on 'DSRC-only' and 'C-V2X-only' taxonomies and have identified the need for more research on coexistence of 'DSRC and C-V2X'. The current national plans for ITS spectrum usage herein examined are similar, but the plans for the future vary and encompass DSRC, C-V2X, and Wi-Fi technologies with Re-CH, interoperability between DSRC and C-V2X without Wi-Fi, or C-V2X only. Even though the regulatory plans differ, the common highest priority is the safety of ITS users.

The objectives of this paper were to provide the current state of the art of ITS spectrum regulation around the world, survey the technical challenges, and explore the emerging challenges where research is critical. This will help canalizing and prioritizing research to enable rapid advances. It will also support transition to practice with support from national spectrum regulation and international standardization while leveraging the diversity of the technical solutions.

This article has been accepted for publication in a future issue of this journal, but has not been fully edited. Content may change prior to final publication. Citation information: DOI 10.1109/ACCESS.2020.3012788, IEEE Access

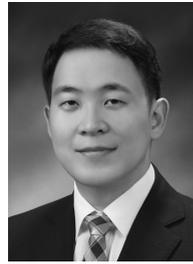

**Junsung Choi** (choijs89@kaist.ac.kr) is currently a researcher in the CCS Graduate school of Green Transportation with KAIST, Daejeon, Korea. He received B.S., M.S., and Ph.D. degree in Electrical and Computer Programming Engineering from Virginia Tech in 2013, 2016, and 2018, respectively. He served as a member of Wireless@VT between 2013 and 2018. During the M.S. and Ph.D. degree, he was a Research Assistant (the Bradley Department of Electrical and Computer Engineering, Virginia Tech). His research interests include vehicular communications, V2X communications, propagation chcannel characteristics, 5G communications, and LPI communications.

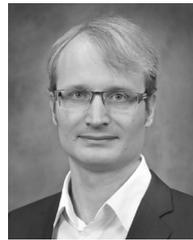

**Vuk Marojevic** (vuk.marojevic@msstate.edu) is an associate professor in electrical and computer engineering at Mississippi State University. He obtained his M.S. from the University of Hannover, Germany, in 2003 and his PhD from Barcelona Tech-UPC, Spain, in 2009, both in electrical engineering. His research interests are in 4G/5G security, spectrum sharing, software radios, testbeds, resource management, and vehicular and aerial communications technologies and systems. He is an Editor of the IEEE Trans. on Vehicular Technology, an Associate Editor of the IEEE Vehicular Technology Magazine, and Officer of the IEEE ComSoc Aerial Communications Emerging Technology Initiative.

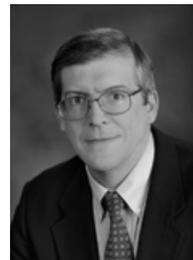

**Carl B. Dietrich** (cdietric@vt.edu) is a Research Associate Professor in the Bradley Department of Electrical and Computer Engineering at Virginia Tech. He earned Ph.D. and M.S. degrees in Electrical Engineering from Virginia Tech, Blacksburg, VA, and a B.S. in Electrical Engineering from Texas A&M University, College Station, TX. His research interests include cognitive radio, software defined radio, multi-antenna systems, and radio wave propagation. Dr. Dietrich has chaired the Wireless Innovation Forum's Educational Special Interest Group, is an IEEE Senior Member and a member of IEEE Eta Kappa Nu and ASEE. He is also a licensed professional engineer in Virginia.

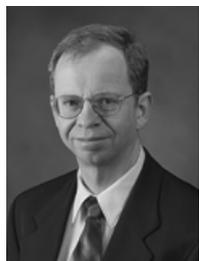

**Jeffrey H. Reed** (reedjh@vt.edu) is currently the Willis G. Worcester Professor with the Bradley Department of Electrical and Computer Engineering, Virginia Tech. He is also the Founder of Wireless @ Virginia Tech, and served as the Director, until 2014. He is also the Founding Faculty Member of the Ted and Karyn Hume Center for National Security and Technology and served as the interim Director when founded, in 2010. He is also a Co-Founder of Cognitive Radio Technologies (CRT), a company commercializing of the cognitive radio technologies; Federated Wireless, a company developing spectrum sharing technologies; and for PFP Cybersecurity, a company specializing in security for embedded systems. His book, Software Radio: A Modern Approach to Radio Design (Prentice Hall) and his latest textbook Cellular Communications: A Comprehensive and Practical Guide (Wiley-IEEE, 2014). In 2005, he became a Fellow to the IEEE for contributions to software radio and communications signal processing and for leadership in engineering education. He is a Past Member CSMAC a group that provides advice to the NTIA on spectrum issues. In 2013, he received the International Achievement Award by the Wireless Innovations Forum. In 2012, he served on the President's Council for the Advisors of Science and Technology Working Group that examine ways to transition federal spectrum for commercial use.

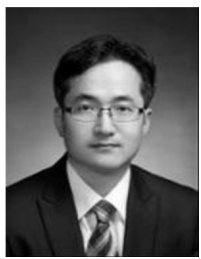

**Seungyoung Ahn** (sahn@kaist.ac.kr) received the B.S., M.S., and Ph.D. degrees from KAIST, Daejeon, South Korea, in 1998, 2000, and 2005, respectively. He is currently an Associate Professor with KAIST. His research interests include wireless power transfer (WPT) system design and electromagnetic compatibility design for electric vehicle and digital systems.